# HEAVY QUARKS ON THE LATTICE

JEFFREY E. MANDULA
DEPARTMENT OF ENERGY
DIVISION OF HIGH ENERGY PHYSICS
WASHINGTON, DC 20585

**Abstract** This lecture describes the treatment of heavy quarks in lattice QCD by implementing the Isgur-Wise limit. The method is briefly discussed, and some of the special features of the resulting theory are highlighted. We emphasize issues of the renormalization of the effective theory. The formulation permits a calculation of heavy quark processes even when the momentum transfers are much larger than the inverse lattice spacing. Applications include semi-leptonic heavy quark decay and scattering processes, including the computation of the nonperturbative part of the Isgur-Wise universal function.

## 1. INTRODUCTION

Heavy quarks are an especially interesting challenge in lattice QCD. In many ways heavy quarks should be much easier to treat than light ones, since their Fermi momentum is typically non-relativistic. However, the three-momentum of even a slow-moving heavy quark can easily be larger than $2\pi/a$ (where $a$ is the lattice spacing), the maximum value of each momentum component that is representable on a lattice. A first glance, then, the whole idea of discussing heavy quarks in a lattice context seems problematical. Fortunately, for some problems we can insure that the quark's momentum is small enough so that it is a good approximation to take a strictly static heavy quark as a starting point, and develop an expansion in either its inverse mass, its velocity, or its momentum. The striking success in calculating the spectrum of $\Upsilon$ resonances, which are approximately $b\bar{b}$ bound states, shows the power of this approach when the momenta are under good control. On the other hand, there are important processes in which one cannot constrain the momenta to be small.

A basic process which necessarily involves dealing with large momenta is the decay of one particle containing a heavy quark into another, such as the semileptonic $B$ meson decay, $B^* \to Dl\nu$. In the continuum there has been great progress in efficiently describing the dynamics of such processes. There is a sensible expansion in the inverse of the heavy quark mass, and for infinite mass, known as the Isgur-Wise limit, notable simplifications obtain. The analysis to be presented here concerns the implementation of this dynamical limit on the lattice. We begin by briefly reviewing the continuum heavy quark theory, and the lattice theory which describes it. We then discuss the renormalization of the resulting Euclidean-space theory, and introduce a method for efficiently performing lattice calculations with it. We conclude with short discussion of some of the more interesting problems that are probably ripe for solution.

## 2. THE ISGUR-WISE LIMIT

The key dynamical idea of the Isgur-Wise limit is that the four-momentum of a heavy quark can be usefully decomposed as $P_\mu = Mv_\mu + p_\mu$, where $v$ is a constant but arbitrary four-velocity normalized to $v^2 = 1$ [1][2][3]. The four-vector $v$ is called the classical velocity of the heavy quark. In the $M \to \infty$ limit, a new symmetry emerges. Part of the symmetry is a flavor $SU(N)$ symmetry between $N$ flavors of heavy quarks. It relates, for example, decays of $B$ mesons to those of $D$ mesons. The more powerful aspect of the symmetry comes

# HEAVY QUARKS ON THE LATTICE

from the fact that in the $M \to \infty$ limit, the spin of the heavy quark becomes dynamically irrelevant. This aspect of the new symmetry relates processes involving particles of different spin, such as $B \to D$ and $B \to D^*$ transitions.

The simplifications of the Isgur-Wise limit are most directly seen in the perturbative fermion propagator. Assume we have a heavy quark of mass $M$, moving with momentum $P$. If we write $P$ as $Mv + p$ and pass to the $M \to \infty$ limit, the fermion propagator simplifies to

$$\frac{1}{\gamma \cdot P - M} \to \frac{1 + \gamma \cdot v}{2 v \cdot p} \tag{1}$$

In the limit the heavy quark mass $M$ and the large momentum $Mv$ disappear, leaving behind only finite terms.

The equivalent to the Isgur-Wise limit in Euclidean coordinate space is the following. The propagator is given by

$$S(x) = \int \frac{d^4 p}{(2\pi)^4} \frac{1}{i\gamma \cdot p + M} e^{ip \cdot x} \tag{2}$$

which satisfies the equation

$$(\gamma \cdot \partial + M) S(x) = \delta(x) \tag{3}$$

By an appropriate shift of contour in the energy plane, one can show that in the $M \to \infty$ limit, for positive Euclidean time,

$$e^{M v_0 x_4 - i M \vec{v} \cdot \vec{x}} S(x) \to \int \frac{d^4 p}{(2\pi)^4} \frac{1 - i\gamma \cdot v}{2 p \cdot v} e^{ip \cdot x} \tag{4}$$

where $v$ is the four-vector $(iv_0, \vec{v})$ and $v_0$ is $[1+\vec{v}^2]^{1/2}$. For $x_4 < 0$, the limiting form of the propagator is the same but with the sign of $v_0$ (and $v_4$) reversed. This represents the heart of the Isgur-Wise limit because there are no kinematically large momenta remaining in the integral. The appearance of the complex energy-momentum four-vector is unusual. It ensures that the propagation of a heavy quark moving with three-momentum $M\vec{v}$ shows a decay in Euclidean time with the correct energy, and that the projection operator appearing in the numerator has the same form it has in Minkowski space.

The heavy quark limit can be compactly expressed in terms of an effective Lagrangian. One defines a reduced field by factoring from the quark field the phase which diverges in the $M \to \infty$ limit,

$$h^{(v)}(x) = \lim_{M \to \infty} e^{-iMv\cdot x}\, \frac{1 + \gamma\cdot v}{2}\, \psi(x) \tag{5}$$

The free field $h^{(v)}$ satisfies a reduced Dirac equation

$$-iv\cdot\partial\, h^{(v)}(x) = 0 \tag{6}$$

and gauge interactions are incorporated simply by replacing the partial with the covariant derivative

$$\partial \to D = \partial + [A,\ldots] \tag{7}$$

For each value of the classical velocity $v$, there is an independent reduced field. The dynamics of $h^{(v)}$ are expressed by the effective Lagrangian

$$\mathcal{L}^{(v)} = \bar{h}^{(v)}(x)\, iv\cdot D\, h^{(v)}(x) \tag{8}$$

Which is known as the Heavy Quark Effective Theory (HQET).

A lattice version of this theory requires an appropriate discretization of the reduced Dirac equation[4]. To implement on the lattice the property that, in the $M \to \infty$ limit, heavy quarks propagate only forward in time drives one to an asymmetric discretization of the derivative operator. The most convenient choice is to use a symmetrical first difference for the spatial derivative but an asymmetric forward difference for the time derivative, a form often referred to as forward-time centered-space (FTCS).

$$v_0 [U(x,x+\hat{t}) S(x+\hat{t},y) - S(x,y)]$$
$$+ \sum_{\mu=1}^{3} \frac{-iv_\mu}{2} [U(x,x+\hat{\mu}) S(x+\hat{\mu},y) - U(x,x-\hat{\mu}) S(x-\hat{\mu},y)] = \delta(x,y) \quad (9)$$

This equation effectively defines the lattice heavy quark theory.

The lattice reduced heavy quark propagator is the solution to this equation. On each lattice in a Monte Carlo ensemble it is obtained by direct forward recursion, which requires only a tiny amount of computation.

If the lattice interaction strength is zero, $[U(x, x+\hat{\mu}) = 1]$, the reduced propagator is proportional to

$$e^{[-E(\vec{k})t + i\vec{k}\cdot\vec{x}]} \quad (10)$$

where the energy E is given by

$$E(\vec{k}) = -\ln\left[1 + \sum_{\mu=1}^{3} \frac{v_\mu}{v_0} \sin(k_\mu)\right] \quad (11)$$

This is the lattice approximation to the continuum behavior

$$E(\vec{k}) = \lim_{M \to \infty} \left[\sqrt{M^2 + (M\vec{v}+\vec{k})^2} - M\sqrt{1+\vec{v}^2}\right]$$
$$= \frac{\vec{v}\cdot\vec{k}}{v_0} \quad (12)$$

## 3. LATTICE HEAVY QUARK RENORMALIZATION

This is surely not the place to review all the standard material on renormalization or even the renormalization of the heavy quark effective theory! The thorough review of Neubert covers the latter, including such subtleties as the matching of the effective heavy quark theory to QCD and the renormalization of composite operators, which enter into the calculation of weak decay rates[5].

On the lattice there is, however, a new renormalization which has no continuum analogue. It is the renormalization of the heavy quark classical velocity. Recall that in introducing the classical velocity, we described it in terms of the division of the heavy quark's momentum into a large fixed (classical) part and a dynamical part whose magnitude is set by the natural energy scale of QCD. We then used this division to arrive at a reduced Dirac equation for the heavy quark field. In principle, these two usages of the heavy quark classical velocity could logically be different.

In the continuum, the values of the velocity that appear in the Dirac equation and as a parameter in the separation of the total four-momentum are always the same. To see this, it is sufficient to realize that $v$ is the only four-vector parameter in the heavy quark theory. Therefore, the classical velocities appearing in those two contexts must be proportional to each other, and since they are by definition both normalized to 1, they are equal.

However, on the lattice this is not the case[6]. A difference is possible because of the reduced symmetry of the lattice relative to continuum space-time. In the lattice heavy quark theory there are many linearly independent "four-vectors" that can be made from the components of a given four-vector $v$, where by a lattice four-vector is meant any quantity that has the same transformation properties as $v$ under the lattice rotation-reflection group. The simplest examples are the vectors $(v_0^{2n+1}, v_1^{2n+1}, v_2^{2n+1}, v_3^{2n+1})$, whose components are the $(2n+1)^{\text{st}}$ power of the corresponding component of $v$.

# HEAVY QUARKS ON THE LATTICE

Let us take as the definition of the bare, or input lattice classical velocity the quantity that appears in the reduced Dirac equation. In perturbation theory, this is the quantity that appears in the bare propagator. The physical classical velocity is the quantity that appears in the break-up of the four-momentum into a part proportional to the heavy quark mass and a part which is finite in the heavy quark limit. On the lattice, the components of this physical classical velocity could be proportional to any linear combination of odd powers of the components of the input classical velocity, subject only to an overall normalization condition. Since the possibility of a renormalization of the classical velocity arises purely because of the discretization of space-time, the actual shift is very sensitive to the details of the how the heavy quark theory is implemented on the lattice.

Renormalization is most frequently discussed in perturbation theory, but in the case of the classical velocity, as we shall see, it is probably not at all reliable. To keep the context clear, when referring to the physical or renormalized classical velocity, I will use different symbols depending on whether it is determined by simulation or by lattice perturbation theory. Throughout this lecture I will use the symbol $v^{(phys)}$ for the nonperturbatively determined physical classical velocity, and the symbol $v^{(ren)}$ for the same quantity when it is calculated in perturbation theory. For the input or bare classical velocity I will use $v^{(input)}$, or simply $v$, except when the perturbative context is to be emphasized, in which case the symbol $v^{(bare)}$ will be used.

In perturbation theory it is most natural to directly evaluate the residual heavy quark propagator, but in simulations is both more natural and more convenient to simulate the residual propagator of a composite particle containing a heavy quark. In order to simulate the heavy quark propagator, we would have to apply a global gauge-fixing procedure, and to exploit residual lattice symmetry we would have to choose a covariant gauge, such as the lattice Landau gauge. In perturbation theory this is all accomplished automatically by the choice of the gluon propagator, but non-perturbatively, gauge fixing would demand rather more computation than the calculation of the heavy quark propagator itself.

JEFFREY E. MANDULA

The physical classical velocity of a meson made of one heavy and one light quark is the same as that of its heavy quark component. The reason for this is because the mass difference between the heavy quark and the composite meson remains finite in the heavy quark limit, but corresponds to a finitely different breakup of the total 4-momentum

$$P = Mv + p \rightarrow (M + m)v' + p' \tag{13}$$

In the heavy quark limit, both $v - v'$ and $p - p'$ vanish like $m/M$, where $m$ is the difference between the composite particle mass and the heavy quark mass $M$.

Let us define the physical classical velocity, $v^{(phys)}$, through the continuum physical residual energy of a particle containing a heavy quark, relative to the energy of its quark component.

$$E^{(v)}(\vec{p}) = \lim_{M \to \infty} \sqrt{(M+m)^2 + ((M+m)\vec{v}^{(phys)} + \vec{p})^2} - M v_0^{(phys)}$$
$$= m v_0^{(phys)} + \boldsymbol{\wp}^{(phys)} \cdot \vec{p} \qquad (\boldsymbol{\wp}^{(phys)} \equiv v_i^{(phys)}/v_0^{(phys)}) \tag{14}$$

From this relation we have

$$\boldsymbol{\wp}^{(phys)} = \left. \frac{\partial E^{(v)}(\vec{p})}{\partial \vec{p}} \right|_{\vec{p} = 0} \tag{15}$$

On an infinite lattice the momentum is continuous (though bounded), so this definition is quite sufficient. On a finite lattice we must specify a finite difference approximation to the momentum derivative. We will use the symmetrical first difference:

$$\boldsymbol{\wp}_i^{(phys)} = \left. \frac{\Delta E^{(v)}(\vec{p})}{\Delta p_i} \right|_{\vec{p} = 0} \equiv \frac{E^{(v)}(p_{min}\hat{\imath}) - E^{(v)}(-p_{min}\hat{\imath})}{2 p_{min}} \tag{16}$$

HEAVY QUARKS ON THE LATTICE

Here $p_{\min} = 2\pi/Na$ is the smallest finite momentum representable on a lattice with spacing $a$ and $N$ sites to a side.

## 4. RENORMALIZATION OF THE CLASSICAL VELOCITY

The simulation of the classical velocity requires the computation of the function $E^{(v)}(\vec{p})$, which controls the asymptotic rate of falloff of a heavy particle's propagator $M^{(v)}(t,\vec{p})$. For fixed $\vec{v}$ and $\vec{p}$, the asymptotic behavior of the propagator is

$$M^{(v)}(t,\vec{p}) \sim C^{(v)}(\vec{p}) \, e^{-E^{(v)}(\vec{p})t} \tag{17}$$

The energy $E^{(v)}(\vec{p})$ is the asymptotic coefficient of the Euclidean time $t$ in the logarithmic derivative of $M^{(v)}$ with respect to $\vec{p}$, evaluated at zero momentum.

$$\begin{aligned}
&\frac{\partial M^{(v)}(t,\vec{p})/\partial p_i \big|_{\vec{p}=0}}{M^{(v)}(t,\vec{p}=0)} \\
&\sim \frac{\partial C^{(v)}(\vec{p})/\partial p_i \big|_{\vec{p}=0}}{C^{(v)}(\vec{p}=0)} - \frac{\partial E^{(v)}(\vec{p})}{\partial p_i}\bigg|_{\vec{p}=0} t \\
&= \frac{\partial C^{(v)}(\vec{p})/\partial p_i \big|_{\vec{p}=0}}{C^{(v)}(\vec{p}=0)} - v_i^{(phys)} \, t
\end{aligned} \tag{18}$$

To adapt this discussion to the actual situation of finite-extent lattices, the only modification required is the replacement of the continuum momentum derivative by a lattice approximation, which we again take to be the symmetrical first difference on the finite Fourier transform lattice.

$$\left.\frac{\partial M^{(v)}(t,\vec{p})}{\partial p_i}\right|_{\vec{p}=0} \Rightarrow \left.\frac{\Delta M^{(v)}(t,\vec{p})}{\Delta i}\right|_{\vec{p}=0}$$
$$\equiv \frac{1}{2p_{min}}\left[M^{(v)}(t,\vec{p}=\vec{\imath}p_{min}) - M^{(v)}(t,\vec{p}=-\vec{\imath}p_{min})\right] \quad (19)$$

### 4.1 Expansion in the Bare Classical Velocity

The structure of the calculation of the propagator $M^{(v)}(t,\vec{p})$ and the quantities derived from it is brought out by expanding in powers of the components of the bare classical velocity:

$$M^{(v)}(t,\vec{p}) = \sum_{m_1,m_2,m_3} v_1^{m_1} v_2^{m_2} v_3^{m_3} M(t,\vec{p},\vec{m}) \quad (20)$$

The gauge invariant propagator $M^{(v)}$ satisfies lattice rotation-inversion symmetry conditions. For the case of a scalar meson, the propagator is invariant under simultaneous lattice transformations of $\vec{v}$ and $\vec{p}$. To express this in terms of the coefficients $M(t,\vec{p},\vec{m})$, recall that the 48-element three-dimensional lattice rotation-reflection group is equivalent to the group of permutations on the three axes times independent inversion groups for each axis[7]. The coefficients are then invariant under simultaneous permutations of the components of $\vec{p} = (p_1, p_2, p_3)$ and $\vec{m} = (m_1, m_2, m_3)$. In addition, $M(t,\vec{p},\vec{m})$ is an even or an odd function of $p_i$ depending on whether $m_i$ is an even or an odd integer, respectively. Note that because of the lattice symmetries, the derivative or symmetrical first difference of $M^{(v)}(t,\vec{p})$ with respect to $p_i$ at $\vec{p} = 0$ must be odd in $v_i$ and even in the orthogonal components $v_j$ $(j \neq i)$.

The basic elements of the simulation are the coefficients $M(t,\vec{p},\vec{m})$ which appear in the expansion of $M^{(v)}$ in powers of the input classical velocity. Since the physical classical velocity is directly extracted from the logarithmic derivative of $M^{(v)}(t,\vec{p})$, we expand it in an analogous series in the input classical momentum $v$:

$$\frac{\Delta M^{(v)}(t,\vec{p})/\Delta p_i}{M^{(v)}(t,\vec{p}=0)} = \sum \mathbf{v}_1^{m_1} \mathbf{v}_2^{m_2} \mathbf{v}_3^{m_3} R^{(i)}(t,\vec{m}) \qquad (21)$$

Each of the expansion functions $R^{(i)}(t,\vec{m})$ is asymptotically linear in $t$, and the negative of its slope is the corresponding term in the expansion for $\mathbf{v}_i^{(phys)}$ in powers of $\mathbf{v}$:

$$R^{(i)}(t,\vec{m}) \sim Z^{(i)}(\vec{m}) - c^{(i)}(\vec{m})\, t$$

$$\mathbf{v}_i^{(phys)} = \sum_{m_1,m_2,m_3} \mathbf{v}_1^{m_1} \mathbf{v}_2^{m_2} \mathbf{v}_3^{m_3}\, c^{(i)}(\vec{m}) \qquad (22)$$

### 4.2 The Heavy Quark Propagator Coefficients

A heavy meson propagator is computed in lattice simulations by calculating on each lattice of an ensemble a heavy quark propagator and a light quark propagator, and averaging the product over the ensemble. The light quark propagator is computed by solving the Dirac equation on the lattice. There are many methods in use, including the original Wilson formulation with a term proportional to the lattice spacing added to control the fermion doubling artifacts that, in one guise or another, must be overcome in all simulations. Since our focus here is those aspects of the simulation that are peculiar to heavy quarks, we will discuss the light quarks no further.

The classical velocity enters through the equation for the reduced heavy quark propagator, Eq. (9). If we divide through the lattice Dirac equation by $v_0$, we see that each component of the three-velocity, $\mathbf{v}_i = v_i/v_0$ plays the role of a transverse hopping constant. That is, the reduced heavy quark propagator gets one factor of $\mathbf{v}_i$ for each unit displacement in the $\pm i^{th}$ direction. Since the propagator at any particular site on the $n^{th}$ time slice depends only on its value at the same spatial site or sites displaced by one transverse lattice spacing on the previous time slice, after n time steps the propagator is an $(n-1)^{st}$ order polynomial in the components $\mathbf{v}_i$. Since the classical velocity enters only through the heavy quark propagator, each coefficient in the series for the meson propagator involves only one term in the heavy quark propagator series, that with just the same value of $\vec{m}$.

The calculation of the heavy quark propagator in simulations is greatly facilitated by exploiting this structure and expanding the propagator in a power series in $v_i$:

$$S^{(v)}(t,\vec{x}) = \sum_{m_1,m_2,m_3} v_1^{m_1} v_2^{m_2} v_3^{m_3} S(t,\vec{x},\vec{m}) \qquad (23)$$

The site $\vec{x}$ is measured from the starting location. The computation of the coefficients in this polynomial is highly efficient. The value of the index $m_i$ is the maximum transverse lattice displacement of the heavy quark propagator in the $i^{\text{th}}$ direction contributing to that coefficient. More precisely, the coefficient $S(t,\vec{x},\vec{m})$ is non-zero only on the sites

$$\begin{aligned} x_i &= -m_i, -m_i+2, \ldots, +m_i \qquad (i = 1,2,3) \\ \sum_{i=1}^{3} m_i &\leq t - 1 \end{aligned} \qquad (24)$$

Since each transverse hop in the computation of $S(t,\vec{x},\vec{m})$ can occur at any time between the initial and final times, the relative growth of these coefficients with $t$ will have a factor

$$S(t,\vec{x},\vec{m}) \propto t^{m_1+m_2+m_3} \qquad (25)$$

### 4.3 Variational optimization of the meson creation operator

It is well known that a gauge-invariant heavy-light meson field consisting of component fields at the same point in space-time is a poor choice in simulations, because many time steps are needed for the excited meson states to die off and for the ground state to dominate its propagator. While any smeared operator containing the reduced heavy quark field can be expected to improve the rate of convergence, in actual simulations of quantities defined via composite fields, the limiting factor in the precision of the final result is usually how quickly the ground state in a given sector dominates the propagator. This is so because the statistical precision of propagators deteriorates rapidly with increasing lattice time, and so it is crucial that the propagator reach its asymptotic form in as few time steps as possible.

Draper, McNeile, and Nenkov[8] have formulated this requirement variationally. Their idea is as follows. In any sector defined by a given set of quantum numbers and a residual momentum, the ground state contribution to a composite field propagator has the slowest rate of decay. Therefore, the requirement that the composite field be chosen so that its propagator reaches its asymptotic form in as few time steps as possible is equivalent to the requirement that the state created by the composite field have as large a component as possible along the ground state. This in turn is equivalent to the requirement that for any given time separation, the composite field is chosen so that the magnitude of its propagator is as large as possible.

Let us consider composite operators that, in the Coulomb gauge, as functions of time and residual momentum have the form

$$\Psi^{(v)}(t,\vec{p}) = \sum_{\vec{y}} \psi^{(v)}_{[\vec{p}]}(\vec{y}) \left[ \sum_{\vec{x}} e^{-i\vec{p}\cdot\vec{x}} q(x) h^{(v)}(x+y) \right] \quad (26)$$

where $q(x)$ is a light quark field, $\psi$ is a relative coordinate weighting function and the sum goes over sites $y$ on the same time slice as $x$. For $\vec{v} = 0$, $\psi$ is effectively the wave function of a composite particle, but in general the weighting function does not have this interpretation. The propagator of this composite field is

$$M^{(v)}_{[\psi]}(t,\vec{p}) = \sum_{\vec{y},\vec{y}'} \psi^{(v)*}_{[\vec{p}]}(\vec{y}) K^{(v)}_{[\vec{p}]}(t,\vec{y},\vec{y}') \psi^{(v)}_{[\vec{p}]}(\vec{y}') \quad (27)$$

where $K$ is the composite field propagator for fixed total three-momentum and arbitrary initial and final separation.

$$K^{(v)}_{[\vec{p}]}(t,\vec{y},\vec{y}\,') = \sum_{\vec{x}} e^{-i\vec{p}\cdot\vec{x}} \left\langle s(x, x'=0) S^{(v)}(x+y, y') \right\rangle \quad (28)$$

The average is over the ensemble of lattices. While it might seem counter-intuitive to take the location of the light quark as nominal coordinate of the meson field, it is by far handiest to do this. The requirement that the normalized propagator be maximal on a given time slice implies that $\psi^{(v)}$ is that eigenvector of $K^{(v)}$ with the largest eigenvalue:

$$\sum_{y'} K^{(v)}_{[p]}(t,y,y')\,\psi^{(v)}_{[p]}(t,y') = \lambda^{(v)}_{[p]}(t)\,\psi^{(v)}_{[p]}(t,y) \quad (29)$$

The largest eigenvalue is, in fact, the value of the optimal meson propagator on time slice $t$.

$$M^{(v)}_{[\psi_{(opt)}]}(t,\vec{p}) = \lambda^{(v)}_{[\vec{p}]}(t) \quad (30)$$

The solution to this eigenvalue problem is facilitated by expanding the kernel, eigenvalue, and eigenvector in power series in the components of the classical velocity $\mathbf{v}_i$.

$$K^{(v)}_{[\vec{p}]}(t,\vec{y},\vec{y}\,') = \sum_{m_1,m_2,m_3} \mathbf{v}_1^{m_1}\mathbf{v}_2^{m_2}\mathbf{v}_3^{m_3}\, K_{[\vec{p}]}(t,\vec{y},\vec{y}\,',\vec{m})$$

$$\lambda^{(v)}_{[\vec{p}]}(t) = \sum_{m_1,m_2,m_3} \mathbf{v}_1^{m_1}\mathbf{v}_2^{m_2}\mathbf{v}_3^{m_3}\, \lambda_{[\vec{p}]}(t,\vec{m}) \quad (31)$$

$$\psi^{(v)}_{[\vec{p}]}(t,\vec{y}) = \sum_{m_1,m_2,m_3} \mathbf{v}_1^{m_1}\mathbf{v}_2^{m_2}\mathbf{v}_3^{m_3}\, \psi_{[\vec{p}]}(t,\vec{y},\vec{m})$$

In terms of these expansion functions, the eigenvector/eigenvalue problem takes a familiar form. It has the structure of a perturbative treatment of an eigenvalue problem, with the components of the input classical velocity playing the mathematical role of coupling strengths. In particular, it is only the static, zeroth order equation that requires finding the eigenvalues of a matrix, and it is only the zeroth order kernel matrix that must be inverted. Both the final time $t$ and the residual momentum of the meson $\vec{p}$ are fixed parameters.

The dependence of the zeroth order kernel on the residual momentum may be explicitly displayed, so that once the $\vec{p} = 0$ problem is solved, the results for non-zero $\vec{p}$ can be immediately read off. The zeroth order kernel

$$K_{[\vec{p}]}(t,\vec{y},\vec{y},\vec{m}=0) = \sum_{\vec{x}} e^{-i\vec{p}\cdot\vec{x}} \langle s(x,x'=0) S(x+y,y,\vec{m}=0) \rangle \quad (32)$$

is quite simple because the residual heavy quark propagator has no transverse hops. Thus the sum over $\vec{x}$ has only one contribution, at $\vec{x} = \vec{y}' - \vec{y}$, which gives the momentum dependence as a diagonal similarity transformation of the zero-momentum kernel:

$$K_{[\vec{p}]}^{(0)}(t,\vec{y},\vec{y}') = e^{i\vec{p}\cdot\vec{y}} \langle s(y'-y,x'=0) S^{(v)}(t,\vec{y}';0,\vec{y}') \rangle e^{-i\vec{p}\cdot\vec{y}'} \quad (33)$$

The zeroth order eigenvalue is thus independent of the residual momentum, as it should be, and the residual momentum dependence of the zeroth order eigenvector is simply a phase.

$$\psi_{[\vec{p}]}(t,\vec{y},\vec{m}=0) = e^{i\vec{p}\cdot\vec{y}} \psi_{[0]}(t,\vec{y},\vec{m}=0)$$
$$\lambda_{[\vec{p}]}(t,\vec{m}=0) = \lambda_{[0]}(t,\vec{m}=0) \quad (34)$$

The non-trivial equation for the $\vec{p} = 0$ zeroth order eigenvalue and eigenvector is:

$$\sum_{\vec{y}'} K_{[0]}(t,\vec{y},\vec{y}',\vec{m}=0) \psi_{[0]}(t,\vec{y}',\vec{m}=0)$$
$$= \lambda_{[0]}(t,\vec{m}=0) \psi_{[0]}(t,\vec{y},\vec{m}=0) \quad (35)$$

The higher terms in the expansion of the eigenvalue equation Eq. (29) are semi-negative-definite inhomogeneous equations. Suppressing the matrix labels ($\vec{y}, \vec{y}\,'$), the equation for the general term is

$$\left[K_{[\vec{p}]}(t,\vec{m}=0) - \lambda_{[\vec{p}]}(t,\vec{m}=0)\right]\psi_{[\vec{p}]}(t,m_i)$$
$$= -\sum_{\vec{m}'}\left[K_{[\vec{p}]}(t,m_i') - \lambda_{[\vec{p}]}(t,m_i')\right]\psi_{[\vec{p}]}(t,m_i - m_i') \quad (36)$$

The sum goes over all $\vec{m}'$ satisfying $0 \le m_i' \le m_i$, but excluding $\vec{m}' = \vec{0}$. These are actually a hierarchy of inhomogeneous equations for the expansion coefficients of both the eigenvalue and the eigenfunction. The equation for the $\vec{m}^{\,th}$ terms involves all the previous ones, that is those with $m_i' \le m_i$, but subject to the strict inequality $\Sigma m'_i < \Sigma m_i$. Note that even though $\psi_{[0]}(t,\vec{y},\vec{m}=0)$ is invariant under cubic transformations of $\vec{y}$, all higher, $\vec{m} \ne \vec{0}$ terms are always asymmetric.

All of the inhomogeneous equations for varying $\vec{m}\,'$ have the same kernel,

$$\left[K_{[\vec{p}]}(t,\vec{y},\vec{y}\,',\vec{m}=0) - \lambda_{[\vec{p}]}(t,\vec{m}=0)\delta_{\vec{y},\vec{y}\,'}\right] =$$
$$e^{i\vec{p}\cdot\vec{y}}\left[K_{[0]}(t,\vec{y},\vec{y}\,',\vec{m}=0) - \lambda_{[0]}(t,\vec{m}=0)\delta_{\vec{y},\vec{y}\,'}\right]e^{-i\vec{p}\cdot\vec{y}\,'} \quad (37)$$

The component of the equation for the general term, Eq. (36) along $\psi_{[\vec{p}]}(t,\vec{y},\vec{m}=0)$ gives the eigenvalue coefficient:

$$\lambda_{[\vec{p}]}(t,\vec{m}) = \sum_{\vec{m}'\ne\vec{0}}^{\vec{m}} \psi^{\dagger}_{[\vec{p}]}(t,\vec{m}=0)K_{[\vec{p}]}(t,\vec{m}')\psi_{[\vec{p}]}(t,\vec{m}-\vec{m}') \quad (38)$$

The kernel restricted to the orthogonal subspace is negative definite and so can be inverted nonsingularly on that space:

$$\psi_{[\vec{p}]}(t,\vec{m}) = -\left\{Q_{[\vec{p}]}\left[K_{[\vec{p}]}(t,\vec{m}=0) - \lambda_{[\vec{p}]}(t,\vec{m}=0)\right]Q_{[\vec{p}]}\right\}^{-1}$$

$$\times \sum_{\vec{m}' \neq \vec{0}}^{\vec{m}} \left[K_{[\vec{p}]}(t,m_i') - \lambda_{[\vec{p}]}(t,m_i')\right]\psi_{[\vec{p}]}(t,m_i - m_i') \quad (39)$$

where

$$Q_{[\vec{p}]}(\vec{y},\vec{y}') = e^{i\vec{p}\cdot\vec{y}}\left[\delta_{\vec{y},\vec{y}'} - \psi_{[0]}(t,\vec{y},\vec{m}=0)\psi_{[0]}^{\dagger}(t,\vec{y}',\vec{m}=0)\right]e^{-i\vec{p}\cdot\vec{y}'} \quad (40)$$

is the projector on the space orthogonal to the ground state. The matrix inverse is nonsingular on the image space of $Q_{[\vec{p}]}$. Note that only a single matrix needs to be inverted throughout this process.

The successive terms in the expansion of the weighting function each have different large Euclidean time behaviors. Specifically, the relative growth with $t$ of the terms in the expansion of the heavy quark propagator with different values of $\vec{m}$ (Eq. (25)) results in the same relative $t$ dependence of the terms in the expansion of the kernel (Eq. (28)). Since $t$ is a fixed parameter in the whole hierarchy of expressions for the wave function components (Eq. (39)), the coefficient $\psi_{[\vec{p}]}(t,\vec{y},\vec{m})$ grows with $t$ like

$$\psi_{[\vec{p}]}(t,\vec{y},\vec{m}) \sim t^{(m_1 + m_2 + m_3)} \quad (41)$$

## 5. SOME SIMULATION RESULTS

This is as far as we can go analytically. To actually solve the variational problem, we must specify over what class of functions we will maximize the normalized propagator Eq. (27), and then numerically solve the resulting matrix eigenvalue problem.

A simulation was carries out on an ensemble of lattices and Wilson light quark propagators created by the Fermilab ACP-MAPS

Collaboration[9]. The ensemble consisted of 42 lattices of size $24^3 \times 48$ with lattice coupling $\beta = 6.1$ along with Wilson quark propagators with hopping constant $\kappa = .154$. Heavy quarks propagate in only one Euclidean time direction, but which direction is conventional. Each lattice therefore provides effectively two independent configurations, one each for association of the direction of time propagation of the heavy quark with the nominal forward or backward direction of lattice time.

**5.1 The Weighting Function**

As the space of functions over which we maximize the composite propagator, we choose those which are non-zero only within a box with $(2N_w + 1)$ sites on a side, and invariant under lattice cubic rotations and reflections within it. We vary $N_W$ to find how sensitive results are to the size of the function space. As a consequence of the symmetry condition, the rank of the square matrix to be inverted is $(N_W + 1)(N_W + 2)(N_W + 3)/6$, which is fortunately much less than $(2N_W + 1)^3$, the value it would take without imposing the requirement that the ground state of the composite particle be a lattice scalar.

Figure 1 shows how the quality of the solution improves with increasing $N_W$. What is plotted is the effective mass of the ground state as a function of time slice, for box sizes ranging from $N_W = 0$ to $N_W = 6$. The effective mass means, in this context, the log of the ratio of the composite propagator on two successive time slices. The statistical errors on each point are determined independently for each value of $N_W$ and each pair of adjacent time slices using a single lattice ensemble of lattices. The curves are from two-exponential fits to the underlying propagators.

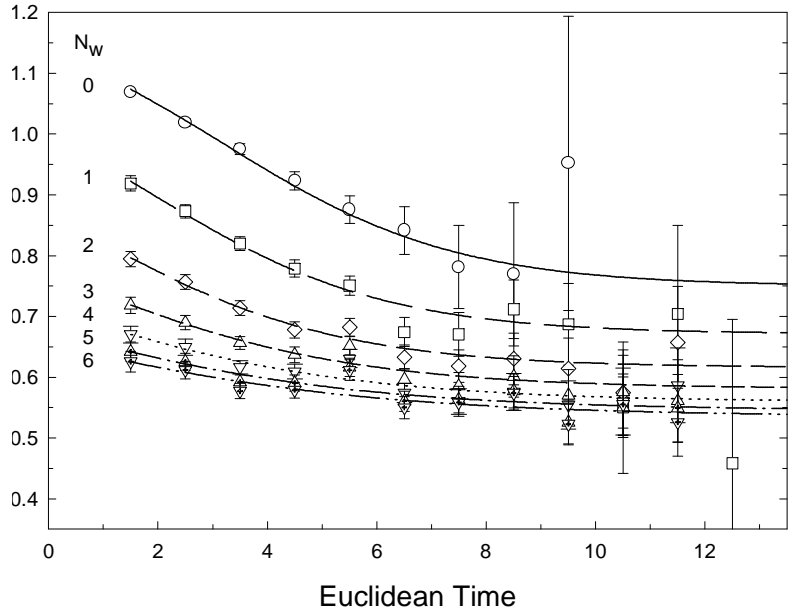

Figure 1 — Convergence to an effective mass plateau of a static heavy-light meson propagator optimized within domains containing $(2N_W + 1)^3$ sites.

There are many lessons to be learned from these plots, which nicely show the trade-offs between finite statistics, which favors short time differences, and the exponential damping of excited state contributions to propagators, which favors long separations, as well as between computing resources, which favors small boxes, and rapid convergence, which favors large ones. For example, it is well known that asymptotically in lattice time only the ground state with given quantum numbers contributes to a propagator with those quantum numbers. This is the source of the statement that any operator with the correct quantum numbers is an acceptable interpolating field. However, it is clear from the two exponential fits that the asymptotic mass seems to depend quite strongly on the size of the box. The reason for this apparent contradiction is that, because of the deterioration of statistics with increasing lattice time, the weighting of successive points in a fit falls off with lattice time. Effectively, only a few, small-separation points actually determine the fit and its attendant asymptotic mass. Since the contribution of the high mass states to the composite propagator at any given lattice time is less for larger boxes, the mass derived from each fit is different. The values derived from each fit converge to the true value as the box size increases. Using larger ensembles of lattices improves statistics, which ameliorates this effect, but can never eliminate it entirely.

With the above remarks as background, let us note several characteristics of the results. The first thing to note is that the local ($N_W = 0$) operator never reaches a plateau. That is, without some sort of smearing procedure for enhancing the convergence of the composite propagator, this program for non-perturbatively computing the renormalization of the classical velocity would fail. Another observation is that the statistical errors, shown by the error bars, are not a true measure of the uncertainty in the values of the effective mass. To see this requires looking closely at the graph. Notice that there are parallel undulations of the data points for the different sized domains relative to the fits. It is the magnitude of these undulations, not the statistics of each data point, that is a meaningful measure of the precision with which the effective mass is determined. The reason that the undulations move in parallel is that they are reflecting correlations from one time slice to the next within the ensemble of lattices. To within that precision, no evident plateau is found for $N_W = 0$ or for $N_W = 1$ (27 sites). For $N_W = 2$ (125 sites), a plateau is reached by about t = 4. For the larger domains, $N_W > 2$ ($7^3$ to $13^3$ sites), a plateau is evident by time slice t = 3. In fact, given the limited precision indicated by the parallel undulations, one cannot be sure that a plateau is not reached even more quickly.

**5.2 The Classical Velocity**

Let us now see how these ideas work in calculating the shift in the classical velocity. Because this shift is a finite renormalization which vanishes in the continuum, as opposed to the familiar situation in which renormalizations result from continuum divergences, it is very interesting to see whether the shift is a large effect or simply a matter of principle without much practical import.

The starting point for the optimization of the composite meson operator described in the previous section is the propagation kernel defined by Eq. (28), or more precisely the coefficients $K_{[\vec{p}]}(t,\vec{y},\vec{y}\,',\vec{m})$ of its expansion in powers of $\vec{v}$. These are the quantities that are the immediate output of a Monte Carlo simulation. We evaluate these coefficients for relative coordinates in a box of with 5 sites on a side ($N_W = 2$). This is a compromise value. The choice $N_W = 3$ would yield a more rapid approach to a plateau, but would require a great deal more computer memory. Higher values of $N_W$ would be prohibitive in their use of computer resources.

The order of battle is as follows. We solve the zeroth order eigenvalue equation, Eq. (35) for its highest eigenvalue and its associated eigenvector with $\vec{p} = 0$. We then use Eq. (34) to obtain the zeroth order eigenvector for $\vec{p} = (1,0,0)$ and its permutations, and then iterate Eqs. (38) and (39). From the classical velocity expansion of

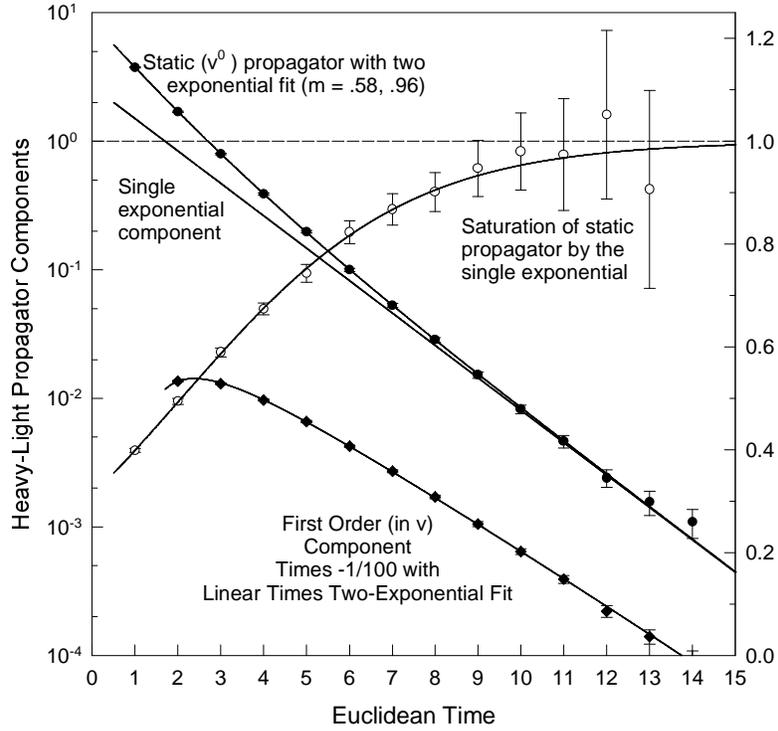

Figure 2 — The zeroth and first order components of the optimized heavy-light meson propagator, showing degree of dominance by the ground state (The right axis is the fractional saturation.)

Eq. (30), or equivalently of Eq. (27), we find the leading term in the classical velocity expansion of the variationally optimized heavy-light meson propagator.

The zeroth and first order terms in the expansion of the heavy-light meson propagator $M^{(v)}$ that enter into the determination of the physical classical velocity are shown in Figure 2. Since terms odd in $m_i$ are odd in $p_i$, the values given for $M(t, \vec{p} = (1..), \vec{m} = (odd..))$ are exactly the same as $p_{min} = 2\pi/Na$ times the lattice approximation to the momentum derivative $\Delta M/\Delta p_i|_{\vec{p}=0}$ of Eq. (19).

There is a lot of information in this graph, and it is worthwhile examining it in a bit of detail. Note the high precision of the leading terms and their convergence to the expected asymptotic forms. The error bars on the computed propagator values are from the single-elimination jackknife analysis, and the error bars on the percentage saturation are the fractional errors on the static propagator. The curve through the static propagator points is a double exponential fit:

$$M(t,\vec{p}=0,\vec{m}=0) \sim c_1 e^{-m_1 t} + c_2 e^{-m_2 t} \quad (42)$$

The masses in the fit are $m_1 = .580 \pm .012$ and $m_2 = .961 \pm .031$, and the ratio of the coefficients is $c_2/c_1 = 2.20 \pm .13$. The higher mass term effectively models all of the higher mass effects in the static propagator. The straight line plotted under the static propagator is the first term from the above fit. The asymptotic mass is, of course, very slightly lower than that obtained from the time-slice to time-slice fall-off the propagator below $t = 10$. The computed saturation values are the single exponential component of the two-exponential fit divided by the propagator values, and the saturation curve is the ratio of the single exponential to that fit.

The curve through the computed values of the first order component of the meson propagator is a fit to the expected asymptotic form of this term, a linear function of lattice time times the two-exponential fit to the static propagator,

$$M(t,\vec{p}=(1,0,0),\vec{m}=(1,0,0)) \sim (a+bt)\left(c_1 e^{-m_1 t} + c_2 e^{-m_2 t}\right) \quad (43)$$

The fits uses the values of the coefficients $c_i$ and $m_i$ determined from the static propagator. As we noted before, the asymptotic slopes of the functions $R^{(i)}(t,\vec{m})$ of Eq. (21) are the negatives of the coefficients of the expansion of the physical classical velocity in powers of the input classical velocity.

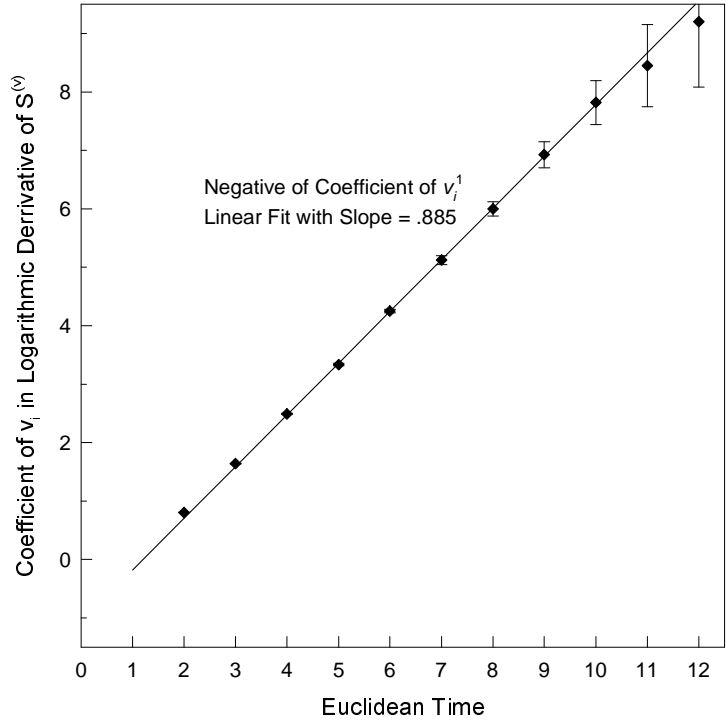

Figure 3 — Leading $v_I$ Expansion Function in the Logarithmic Derivative of the Meson Propagator

Figure 3 is the graph of the negative of the leading function, $R^{(i)}(t, m_i = 1, m_{j \neq i} = 0)$. The indicated statistical errors are obtained from a single-elimination jackknife procedure. This leading term in the logarithmic derivative of the meson propagator apparently achieves its "asymptotic" slope on the very first time slice on which it is non-zero, $t = 2$. The statistical error in the fitted slope, .885, is $\pm$ .017, and so to leading order the shift is $v_i^{(phys)}/v_i^{(input)} = .885 \pm .017$. Corrections come from the cubic and higher powers of $v_i^{(input)}$ in the expansion of $v_i^{(phys)}$.

It is both interesting and disturbing to compare this result to a perturbative, one-loop calculation of the renormalization of the classical velocity[6]. For the lattices used in the simulation, the coupling constant is $g^2 = 6/\beta = 6/6.1$, and the linear term in the shift of $v_i$ is -.23345569. The comparison of the leading terms is

$$v_i^{(phys)}/v_i^{(input)} = \begin{matrix} .885 \pm .017 & simulation \\ .76654431 & one\ loop \end{matrix} \qquad (44)$$

The disagreement between the two methods is quite substantial, because the perturbative calculation evaluates the difference of the above ratio from 1. There are, of course, higher loop terms in perturbation theory, but to bring these two results into concord, higher loops would need to cancel just about half of the leading term.

This is by no means the first example of a serious discrepancy between lattice simulation and lattice perturbation theory. However, the rescaling of the classical velocity is different from other lattice/continuum matching processes is that it seems to be exacerbated rather than improved by the "Tadpole Improvement" procedure of Lepage and Mackenzie[10].

A principal quantitative effect of Lepage and Mackenzie's analysis to introduce a scaling between the couplings used in simulations with those used in the perturbative analysis. It arises because on the lattice there are tadpole corrections to the gauge field, and to cancel these lattice artifacts, one replaces each link variable $U_\mu(n)$ by that link divided by its average value, $U_\mu(n)/u_0$, where the average value of the link is taken to be the gauge-invariant expression

$$u_0 = \left[\frac{1}{3}\langle Tr\,\Box\rangle\right]^{1/4} \qquad (45)$$

and □ denotes the product of the link variables around an elementary 1×1 plaquette. For the Wilson action used to generate the Fermilab lattices, this effects the change

$$I = -\frac{6}{g^2} \sum_\square \frac{1}{3} Tr\,\square \;\Rightarrow\; -\frac{6}{g^2 u_0^4} \sum_\square \frac{1}{3} Tr\,\square \qquad (46)$$

and so $\beta$ is identified with $6/g^2 u_0^4$ rather than with $6/g^2$. Since $u_0$ is necessarily less than one, this always has the effect of increasing the value of the effective perturbative coupling corresponding to a lattice simulation at a given value of $\beta$, thereby increasing the discrepancy of the perturbative estimate of the classical velocity shift from the result of the simulation. Perhaps the fact that tadpole "improvement" worsens the disagreement between the two calculations is related to the fact that the classical velocity renormalization has no divergent part and vanishes in the continuum, but this is just conjecture.

## 6. THE FUTURE – THE HEAVY QUARK FORM FACTOR

The paradigmatic heavy quark process is the weak decay of one meson containing a heavy quark into another, such as the $B$ meson decays

$$B \rightarrow D\,l\,\nu \;,\; D^*l\,\nu$$
$$\langle D v' | \bar{c}\, \gamma_\mu (1 - i\gamma_5) b | B v \rangle \;\sim\; \xi(v\cdot v') \qquad (47)$$

All the dynamical information is contained in the function $\xi$, known as the Isgur-Wise function, or the heavy quark universal form factor. It is a function only of the dot product of the classical velocities, and is normalized to unity for forward decay ($\xi(1) = 1$) because the same form factor appears in the $B$ meson matrix elements of the conserved vector current. The amplitude is contained in the reduced three-point function

$$G^{(v_B,v_D)}(x_0,z_0,y_0) = \left\langle Tr\, s(y,x)\Gamma_D\, S^{(v_D)}(x,z)\Gamma_J\, S^{(v_B)}(z,y)\Gamma_B \right\rangle \quad (48)$$

which is schematically illustrated in Figure 4.

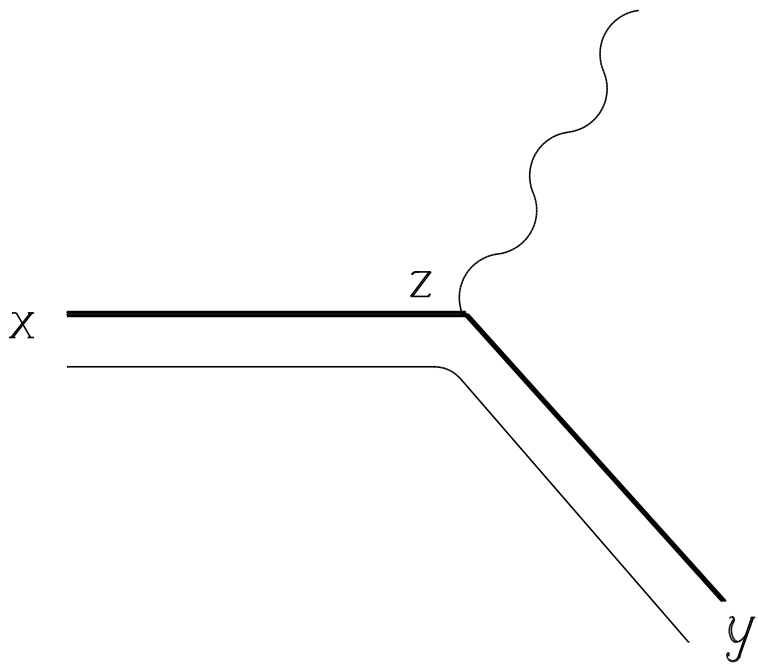

Figure 4 — A heavy meson decay in which one heavy quark emits a weak current and transforms into a different heavy quark

The Isgur-Wise function may be extracted from the three-point correlation by dividing out the normalized single meson propagators. A better way is to take advantage of the normalization in the forward direction and use the three-point function at $v_D = v_B$ as normalization, giving

$$|\xi(v \cdot v')|^2 = \lim_{x_0 \gg z_0 \gg y_0} \frac{G^{(v,v')}(x_0,y_0,z_0) \; G^{(v',v)}(x_0,y_0,z_0)}{G^{(v,v)}(x_0,y_0,z_0) \; G^{(v',v')}(x_0,y_0,z_0)} \qquad (49)$$

This form has the advantage that the overall normalization of the three-point function drops out completely. Only that part of the renormalization that is classical-velocity-dependent need be included. Unfortunately, this renormalization is not at all simple to compute by simulation, but it is crucial to deriving physical results from this analysis. The reason is that the slope of the Isgur-Wise function at the kinematic origin is needed in phenomenological applications. This is a key next step in the study of the lattice heavy quark theory.

The renormalization of the classical velocity affects the calculation of the slope of the Isgur-Wise function using the lattice HQET simply because it changes the heavy quark momentum scale. Of course, only the linear, multiplicative shift enters into the slope of the Isgur-Wise form factor. The cubic and higher order terms that we have not calculated only affect its higher derivatives. The precise effect is seen by expressing the standard kinematic variable $w = v \cdot v' - 1$ in terms of the bounded classical velocity $\wp$

$$w = v \cdot v' - 1 = \frac{1}{2}(\wp - \wp')^2 + O((\wp,\wp')^4) \qquad (50)$$

We see that it is second order in $\wp$. Thus the rescaling $\wp \to \wp^{(phys)}$ has the effect of increasing the computed slope of the form factor with respect to $v \cdot v'$ at the origin by the square of the first order rescaling coefficient,

$$\frac{1}{c^{(i)}(m_i = 1)^2} \approx 1.28 \pm .05 \qquad (51)$$

## 7. CONCLUDING REMARKS

Let us sum up what we haave learned about the heavy quark effective theory on the lattice. We showed how to formulate the Isgur-Wise limit in Euclidean space, and examined some of the properties of the a lattice discretization of the reduced Dirac equation for the residual heavy quark propagator. We examined in detail the renormalization of the classical velocity in this treatment. The origin of this renormalization, which is purely a lattice effect and is not present in the continuum, is the reduction of Lorentz invariance to hypercubic symmetry on the lattice. We reviewed the result of a lattice simulation of the first-order, multiplicative shift, and obtained

$$v_i^{(phys)} = (.885 \pm .017)\, v_i + O(v_i^3, v_i v_j^2)$$

We discussed how the lattice HQET provides a method for calculating the nonperturbative functions that arise in the heavy quark limit of QCD, especially the Isgur-Wise form factor, but noted that one must first carry out a simulation of the renormalization of the composite weak current operator. Finally, we noted that the renormalization of the classical velocity will have a noticeable effect in such calculations by effectively shifting the scale of the Isgur-Wise kinematic variable.

## ACKNOWLEDGMENTS

It is a pleasure to acknowledge the contributions of my collaborator Professor Michael Ogilvie to the work described here. I would also like to express my thanks to Professor T. D. Lee, who provided me the opportunity to present this work at this "Probing Matter" Symposium. This work has been supported by the United States Department of Energy.